# The Golden Mean and the Physics of Aesthetics

Subhash Kak
Louisiana State University, Baton Rouge

**Abstract**

The golden mean, Φ, has been applied in diverse situations in art, architecture and music, and although some have claimed that it represents a basic aesthetic proportion, others have argued that it is only one of a large number of such ratios. We review its early history, especially its relationship to the Mount Meru of Pingala. We examine the successive divisions of 3, 7, and 22 notes of Indian music and suggest derivation from variants of the Mount Meru. We also speculate on the neurophysiological basis behind the sense that the golden mean is a pleasant proportion.

*Dedicated to the memory of José Maceda (1917-2004)*

# Introduction

Two and a half years ago, in 2002, the celebrated Philippine composer José Maceda organized a Symposium on "A Search in Asia for a New Theory of Music" in Manila. His objective was to explore Asian mathematical sources that may be helpful in providing a direction to world music everywhere. The papers at the Symposium have since appeared as a book (Buenconsejo, 2003). At this meeting, Maceda had many conversations with me regarding what he called the "aesthetics of Asian court musics which seek permanence with little change." I told Maceda that I considered the *dhvani*-approach to aesthetics particularly insightful, since it made the audience central to the process of illuminating the essential sentiment behind the artistic creation (e.g. Ingalls *et al*, 1990). According to this view the permanence being sought is the "universal" that can only be approached and never quite reached by the diverse paths that represent different cultural experiences.

Maceda appreciated the fact that physiological geometry plays in perception and aesthetics, but he was emphatic that there was a cultural component to it and that there could be no unique canon of beauty in art. I take this thought as the starting point of my discussion of the golden proportion (Φ = 1.618..), defined by Φ = 1/(Φ – 1), which some theorists have claimed is fundamental to aesthetics. The conception of it as a pleasing shape arose in the late 1800s and there are no written texts that support such usage in the ancient world. Indeed, many non-Western cultures either do not speak of a unique ideal or consider other ratios as ideal, such as √2 of the Islamic architecture.

Adolf Zeising (1854) appears to have been the first to propose the romantic idea that in the golden ratio "is contained the fundamental principle of all formation striving to beauty and totality in the realm of nature and in the field of the pictorial arts." Two prominent 20th century architects, Ernst Neufert and Le Corbusier, used this ratio deliberately in their designs. More recently, many industrial designs (such as the dimensions of the ubiquitous credit card) have been based on it. *Avant-garde* musicians such as Debussy, Bartok, and Xenakis have composed melodic lines where the intervals are according to this ratio. Salvador Dali used this proportion in some canvases. Cartwright et al (2002) argue its importance in music.

But such usage does not make it an aesthetic law, and Mario Livio (2002a) debunks this pretension effectively, showing that the research on which this claim has been made is ambiguous at best. I agree with Livio's warning that "literature is bursting with false claims and misconceptions about the appearance of the Golden Ratio in the arts (e.g. in the works of Giotto, Seurat, Mondrian). The history of art has nevertheless shown that artists who have produced works of truly lasting value are precisely those who have departed from any formal canon for aesthetics. In spite of the Golden Ratio's truly amazing mathematical properties, and its propensity to pop up where least expected in natural phenomena, I believe that we should abandon its application as some sort of universal standard for 'beauty,' either in the human face or in the arts." (Livio, 2002b)

In this note I first present the background to the golden mean and its relationship to the Mount Meru (*Meru-Prastāra*) of Pingala (c. 200 BC but perhaps 450 BC if the many textual notices about him being the younger brother of the great grammarian Panini are right) and its subsequent exposition by other mathematicians. The term "golden section" (*goldene Schnitt*) seems to first have been used by Martin Ohm in the 1835 edition of the textbook *Die Reine Elementar-Mathematik*. The first known use of this term in English is in James Sulley's 1875 article on aesthetics in the 9th edition of the *Encyclopedia Britannica.*

Next, I present multiplicative variants of Mount Meru and the Fibonacci sequence that may explain why the octave of Indian music has 22 micronotes (*śruti*), a question that interested Jose Maceda. I also discuss other sequences related to the Mount Meru, including those that were described by the musicologist and musical scale theorist Ervin Wilson. I conclude with a discussion of the question of aesthetic universals that was raised by José Maceda.

## Historical Background

As historical background, one must go to Pingala's *Chandahśāstra* 8.32-33 (see, for example, Weber, 1863; Singh, 1936; Singh, 1985; Nooten, 1993), in which while classifying poetic meters of long and short syllables, Pingala presents the Mount Meru (*Meru-Prastāra,* 'steps of Mount Meru', but *Meru-Khanda*, 'portion of Mount Meru' by Bhāskara II in his *Līlāvatī* written in 1150), which is also known as the *Pascal's triangle* (Figure 1). The shallow diagonals of the Mount Meru sum to the Fibonacci series, whose limiting ratio is the golden mean (Figure 2). Pingala's cryptic rules were explained by later commentators such as Kedāra (7[th] century) and Halāyudha (10[th] century).

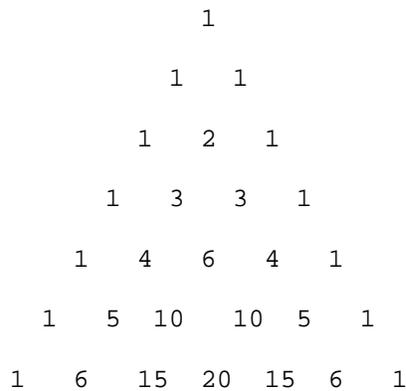

```
                    1
                 1     1
              1     2     1
           1     3     3     1
        1     4     6     4     1
     1     5    10    10     5     1
  1     6    15    20    15     6     1
```

**Figure 1:** Mount Meru

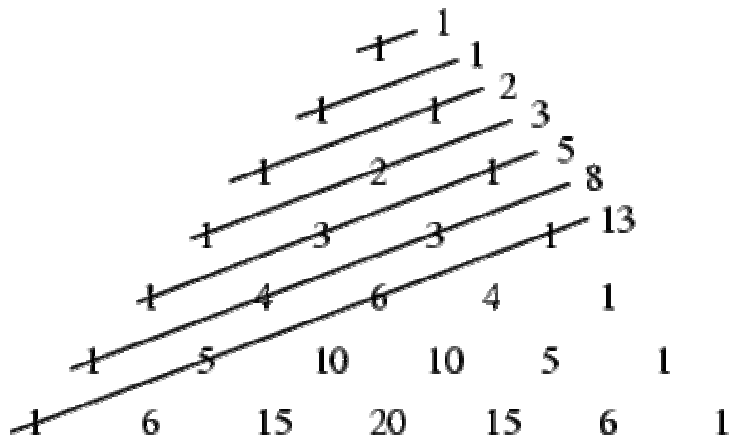

**Figure 2:** Fibonacci Series from Mount Meru

The Fibonacci numbers are described by several Indian mathematicians in the centuries following Pingala as being produced by the rule *F(n+1) = F(n) + F(n-1)*. Virahānka (7[th] century) explicitly gives the sequence 3, 5, 8, 13, 21. Further explanation regarding the numbers was presented by Gopāla (c. 1135) and the polymath Hemacandra Sūrī (1089-1172). If we assume that long syllables take twice as long to say as short ones, Hemachandra (to use the more common spelling of his name) was dealing with the problem: "If each line takes the same amount, what combination of short (S) and long (L) syllables can it have?" The general answer is that a line that take n time units can be formed in F(n) ways. This was explicitly stated by Hemachandra in about 1150 AD.

In Europe, Fibonacci's *Liber Abaci* in 1202 describe these numbers; the book was meant to introduce the Indian number system and its mathematics which he had learnt in North Africa from Arab teachers while a young man growing up there. Fibonacci speaks of his education in North Africa thus: "My father, who had been appointed by his country as public notary in the customs at Bugia acting for the Pisan merchants going there, summoned me to him while I was still a child, and having an eye to usefulness and future convenience, desired me to stay there and receive instruction in the school of accounting. There, when I had been introduced to the art of the Indians' nine symbols through remarkable teaching, knowledge of the art very soon pleased me above all else and I came to understand it."

It has been suggested that the name Gopala-Hemachandra numbers be used for the general sequence:

    *a, b, a+b, a+2b, 2a+3b, 3a+5b, ...*

for any pair *a, b,* which for the case *a=1, b=1* represents the Fibonacci numbers. This would then include the Lucas series for which *a=2 and b=1*. It would also include other series such as the one for which *a=1* and *b=21*, which generates the numbers:

    *1, 21, 22, 43, 65, 108, ….*

Narayana Pandita's *Ganita Kaumudi* (1356) studies additive sequences where each term is the sum of the last *q* terms. He poses the problem thus: "A cow gives birth to a calf every year. The calves become young and they begin giving birth to calves when they are three years old. Tell me, O learned man, the number of progeny produced during twenty years by one cow." (See Singh, 1936; and Singh, 1985 for details)

The Mount Meru was also known in other countries. The Chinese call it "Yang Hui's triangle" after Yang Hui (c. 1238-1298), and the Italians as "Tartaglia's Triangle," after the Italian algebraist Tartaglia, born Niccolo Fontana (1499-1557), who lived a century before Pascal. In Western European mathematical literature, it is called Pascal's triangle after the *Traité du triangle arithmétique* (1655) by Blaise Pascal, in which several results then known about the triangle were used to solve problems in probability theory. Interesting historical anecdotes regarding the combinatoric aspects of these numbers are given in Section 7.2.1.7 of the classic *The Art of Computer Programming* of Knuth (2004).

Here we use the name Mount Meru since it has also had fairly wide currency in the English literature (see, e.g. McClain, 1976), and because it was used by Ervin Wilson whose work will be discussed here. Wilson describes recurrence sequences that he calls Meru 1 through 9.

## A Multiplicative Mount Meru and a Multiplicative Sequence of Notes

The problem of why Indian musicologists speak of a sequence of 3, 7, and 22 notes has been of long-standing interest (see, e.g., Clough *et al*, 1993; Kak, 2003). Although Clough *et al* (1993) present it as the problem of the passage from a chromatic universe of 22 divisions to a "diatonic" set of seven degrees, there is no textual evidence in support of a prior conception of 22 micronotes. For example, Bharata Muni's *Nātya Śāstra* speaks both of the 7 and the 22 divisions.

We propose that a combinatoric consideration may have been behind the choice. Given that the study of poetic meters was an important part of education at that time, it is likely that variants of Mount Meru were examined. An interesting variant on the standard Mount Meru is a Multiplicative Meru where each descendent number is a product of the numbers above it plus one, with the further condition that when the multiplicands are both 0, the product remains 0. One can see in Figure 3 that all other numbers are 0s, and therefore the product at the edges is $(0 \times 1 + 1 = 1)$.

```
                    1
                 1     1
              1     2     1
           1     3     3     1
        1     4     10    4     1
     1     5     41    41    5     1
  1     6     206   1682  206   6     1
```

**Figure 3:** A Multiplicative Mount Meru

An interesting sequence emerging from a similar multiplicative logic is $M(n+1) = M(n) \times M(n-1) + 1$, which gives us:

   *0, 1, 1, 2, 3, 7, 22, 155, 3411, 528706, ....*

We propose that the occurrence of the numbers 3, 7, and 22, in this series is not a coincidence and the above construction was behind the identification of the 22 micronotes in the Indian theory of music. The original number is taken to be 3 basic notes that become the standard 7 notes and the 22 micronotes (*śruti*). Although there is no way to confirm that this was indeed the case, it appears plausible.

## General Recurrence Sequences

In general, we can write the sequence using the linear recurrence relation:

$$F(n) = \sum_{i=1}^{K} a(i)\, F(n-i)$$

where *a(i)* values represent suitable constants. The limiting Ratio, in this case, will be a solution to the algebraic equation:

$$x^K - a(1)\, x^{K-1} - a(2) x^{K-2} - \ldots a(K) = 0$$

It is obvious that any number of Ratios can be obtained by a proper choice of *K* and *a(i)*s. For the case where the *a(i)*s are each 1, the characteristic equation for the Ratio may be simplified to:

$$x^{K+1} - 2\, x^K + 1 = 0$$

Less radical variants of Mount Meru and that of Fibonacci sequences may also be conceived. For example, Wilson's variations of Fibonacci sequences (Wilson 1993, 2001) are obtained by using different slopes of Mount Meru by combining two earlier terms in the sequence. They were named by him as Meru 1 through Meru 9.

**Wilson's Meru 1 through Meru 9**

1. Meru 1: $A_n = A_{n-1} + A_{n-2}$ with limiting ratio 1.618033.. (golden mean)
2. Meru 2: $B_n = B_{n-1} + B_{n-3}$ with limiting ratio 1.465571..
3. Meru 3: $C_n = C_{n-2} + C_{n-3}$ with limiting ratio 1.324717..
4. Meru 4: $D_n = D_{n-1} + D_{n-4}$ with limiting ratio 1.380277..
5. Meru 5: $E_n = E_{n-3} + E_{n-4}$ with limiting ratio 1.220744..
6. Meru 6: $F_n = F_{n-1} + F_{n-5}$ with limiting ratio the same as Meru 3.
7. Meru 7: $G_n = G_{n-2} + G_{n-5}$ with limiting ratio 1.236505..
8. Meru 8: $H_n = H_{n-3} + H_{n-5}$ with limiting ratio 1.193859..
9. Meru 9: $I_n = I_{n-4} + I_{n-5}$ with limiting ratio 1.167303..

The explicit use of these exotic sequences to create musical scales has been presented by the composer Warren Burt (2002). His *Wilson Installations*, a set of three live solo electronic music performances, dealt "with the notions of extended tuning systems, the relation of tuning to timbre and spatiality of sound, and the concentration of attention for extended time periods."

One may also generalize the Mount Meru structure. In Figure 4 we show a triplicate Mount Meru, for which one of the characteristic sequences would be *F(n+1) = F(n) +*

*F(n-1) + F(n-2),* that is, *1, 1, 2, 4, 7, 13, 24, 44, 81, ...* Another example is the sequence *2, 2, 3, 7, 12, 22, 41, 75, 138, ...* that generates not only 3, 7, and 22, but also 12, another characteristic number of Indian music.

```
                    1
                 1  1  1
              1  2  3  2  1
           1  3  6  7  6  3  1
        1  4 10 16 19 16 10  4  1
     1  5 15 30 45 51 45 30 15  5  1
```

**Figure 4:** A Triplicate Mount Meru

## Structural Considerations

If a golden rectangle is drawn and a square is removed, the remaining rectangle is also a golden rectangle (Figure 5a). Continuing this process and drawing circular arcs, the curve formed approximates the logarithmic spiral, which is a form found in nature (Figure 5b). These numbers are encountered in many plants in the arrangement of leaves around the stem, pine cones, seed head packing, and flower petals owing to optimality conditions.

Since physical reasons underlie this occurrence of the golden mean in nature, it is tempting to look at the structural basis of why it is a pleasing proportion cognitively, although we must remember it is not the only one.

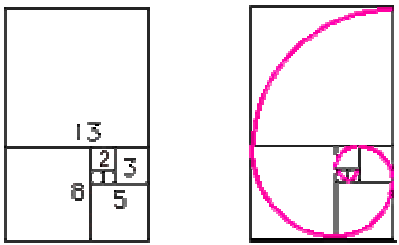

**Figure 5:** The Fibonacci sequence in (a) two dimensions and (b) as a spiral

If the stimulation inside the brain corresponding to an input spreads through a spiral function, then the pleasantness of the golden mean would be clear. On the other hand, neurophysiological structures are not quite two-dimensional, and therefore, there would be a component of the spiral that would go into the third dimension, providing a departure from the "normative" golden mean to many other similar numbers, such as Meru 2 through 9 of Wilson, or other more general numbers. Since the details of the structures are unique to the individual, there is further variability regarding what would be optimal. Nevertheless, inanimate, living and cognitive systems show similar behaviour, which is a consequence of pervasive recursionism (Kak, 2004).

## Concluding Remarks

This note has reviewed some variants of Fibonacci sequences that are of interest to musicologists. The Fibonacci sequence corresponds to a spreading function that is two-dimensional:

$$x^2 = x + 1$$

that is, with the next step as the square of the previous one (Figure 4a). By analogy, in a purely three-dimensional spreading function, the operative formula should be:

$$x^3 = x + 1$$

which corresponds to the recurrence relation of Meru 3. This may be written, alternatively, as the solution of the balanced-looking equation $\Psi = 1/[(\Psi - 1)(\Psi + 1)]$. Since the neurophysiological cognitive structures do not have a symmetry across all the three dimensions (there is also variation across individuals), proportions intermediate to the golden mean and Meru 3 (1.324717…) would also be aesthetically pleasing, especially in traditions which are contemplative, as in Asia.

Coming back to the question of aesthetic universals that Maceda posed, I agree with him that it is most likely that they do not exist. It is cultural authority and tradition that creates them, although they may be shaped by "universals" associated with our cognitive systems.